\documentclass[twocolumn,showpacs,preprintnumbers,amsmath,amssymb]{revtex4}
\usepackage{graphicx}% Include figure files
\usepackage{dcolumn}% Align table columns on decimal point
\usepackage{bm}% bold math

\begin{document}

%\preprint{Phys. Rev. D \textbf{76}, 124004 (2007)}

\title{Revisit on ``Ruling out chaos in compact binary systems"}% Force line breaks with \\

\author{Xin Wu$^{1}$}
\email{xwu@ncu.edu.cn}
\author{Yi Xie$^2$}
\affiliation{1. Department of Physics, Nanchang University,
Nanchang 330031, China \\ 2. Department of Astronomy, Nanjing
University, Nanjing 210093, China}


\begin{abstract}
Full general relativity requires that chaos indicators should be
\emph{invariant} in various spacetime coordinate systems for a
given relativistic dynamical problem. On the basis of this point,
we calculate the \emph{invariant} Lyapunov exponents (LEs) for one
of spinning compact binaries in the conservative second
post-Newtonian (2PN) Lagrangian formulation without the
dissipative effects of gravitational radiation, using the
two-nearby-orbits method with projection operations and with
coordinate time as an independent variable. It is found that the
actual source leading to zero LEs in one paper but to positive LEs
in the other does not mainly depend on rescaling, but is due to
two slightly different treatments of the LEs. It takes much more
CPU time to obtain the stabilizing limit values as reliable values
of LEs for the former than to get the slopes (equal to LEs) of the
fit lines for the latter. Due to coalescence of some of black
holes, the LEs from the former are not an adaptive indicator of
chaos for comparable mass compact binaries. In this case, the
\emph{invariant} fast Lyapunov indicator (FLI) of two nearby
orbits, as a very sensitive tool to distinguish chaos from order,
is worth recommending. As a result, we do again find chaos in the
2PN approximation through different ratios of FLIs varying with
time. Chaos cannot indeed be ruled out in real binaries.
\end{abstract}

% Nonlinear dynamics and chaos, Chaotic dynamics,  Galactic nuclei (including black holes),
% circumnuclear matter, and bulges
\maketitle

The chaotic behavior of nonlinear dynamical systems has become a
very interesting subject in relativistic astrophysics [1].
Especially the dynamics of binary systems of spinning compact
objects in the frame of general relativity does deservedly receive a
great deal of attention. Merging binaries are regarded as the most
promising candidates for future ground- and space-based
gravitational wave detectors, such as LIGO [2]. The successful
detection is necessary to rely on the matched filtering technique
with the theoretical gravitational wave templates matched to
experimental data containing a lot of instrumental noise. However,
chaos in the gravitational wave sources would affect the treatment
of observational data, for example, signals not to be drawn out of
the noise. For this reason, there have been a series of articles
[3-9] for discussing whether spinning compact binaries can exhibit
chaos. In the light of the method of fractal basin boundaries, an
earlier paper [3] emphasized that the contribution of spin-orbit
(SO) and spin-spin (SS) coupling is in favor of chaos for the case
of comparable mass compact binaries in the Lagrangian formulation to
2PN order with the dissipative effects of gravitational radiation
turned off. While another work [4] suggested ruling out chaos by
finding no positive LEs of trajectories  along the fractal basin
boundaries of Ref. [3]. At once, as an answer to this claim, it was
reported in Refs. [5,6] that the wrong results of Ref. [4] should be
owing to the less rigorous calculation of the LEs of two nearby
orbits, with unapt renormalization time steps adopted. Further some
orbits with positive LEs were given.

Obviously, it is very surprise that Ref. [4] and Refs. [5,6]
employed the same chaos index---the largest LE, but gave
completely different dynamics to the same 2PN equations of motion
for spinning compact binaries. Although the latter pointed out the
problem of the former, such interpretation seems still to be
ambiguous and puzzling in retrospect. Here are more thorough
comments on these works. LEs, as a common chaos indicator, measure
the rate of exponential divergence between neighboring
trajectories in the phase space. There are two different methods
to compute them. Historically, the tangent vectors
$\mbox{\boldmath$\xi$}(0)$ and $\mbox{\boldmath$\xi$}(t)$ about a
given trajectory at times 0 and $t$ are used to define the maximum
LE: $ \lambda = \lim_{t\rightarrow\infty}\chi(t)$, with
 $\chi(t) = (1/t)\ln[|\mbox{\boldmath$\xi$}(t)|/|\mbox{\boldmath$\xi$}(0)|].$
The technique for getting the LE is called as method 1 (M1).
Usually it is a cumbersome task to derive the variational
equations associated to the tangent vector for complicated
dynamical systems. For an alternative procedure to M1, a simpler
way, M2, is to adopt the distance $d(t)$ in the phase space
between two nearby trajectories as an approximation to the norm of
the tangent vector $\mbox{\boldmath$\xi$}(t)$ such that
$\chi(t)=(1/t)\ln [d(t)/d(0)]. $ It is for a suitable choice of
the starting separation $d(0)$ and of the rescaling interval that
M2 gives almost the same values of LEs as M1 does (for details,
see Ref. [10]). As a practical application of M2, traditionally
one plots a curve of $\ln \chi(t)$ vs $\ln t$. A negative constant
slope of the curve means the regularity of the system. If the
slope tends gradually to zero and $\ln \chi(t)$ reaches nearly a
stabilizing value, the bounded system becomes chaotic. The diagram
method is marked as M2a. In addition, there is another diagram
method (M2b) by plotting $\ln [d(t)/d(0)]$ vs $t$. It is vital to
perform a least-squares fit on the simulation data to work out the
slope $\chi$ of the fit line $\ln [d(t)/d(0)]=\chi t$, as the
largest LE. There should have been no difference between M2b and
M2a in principle, but the former superior to the latter lies in
that it is much easier to identify the linear growth of $\ln
[d(t)/d(0)]$ than to identify the convergence of $\ln \chi(t)$. In
other words, generally it costs a rather long time for $\ln
\chi(t)$ to converge a limit value in the chaotic case. As to the
fit slope, perhaps it is not very true but can always easily be
seen even if time is short. In particular, the difference is
rather explicit when the authors of Ref. [4] and those of Refs.
[5,6] used M2a and M2b to treat the LEs of compact binaries,
respectively. Thus we think the very slow convergence of $\ln
\chi(t)$ leading to the ``false" LEs in Ref. [4], but do not agree
with the authors of Refs. [5,6], who claimed that the wrong LEs in
Ref. [4] depend on the rescaling. This will be further checked in
our next numerical experiments, where both M2a and M2b adopt the
same rescaling interval.

\begin{figure}[h]
\includegraphics[scale=0.85]{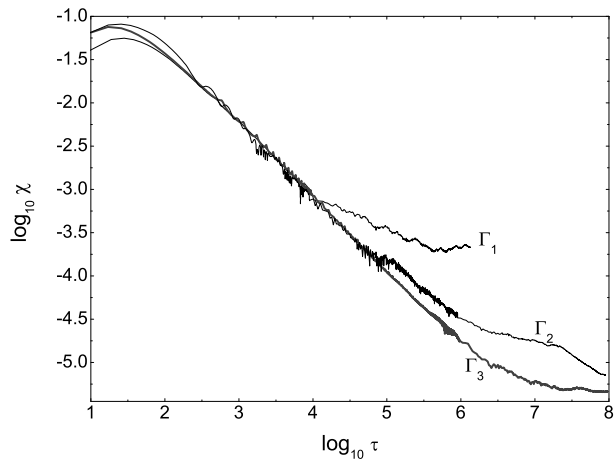}% Here is how to import EPS art
\caption{Invariant LEs of the considered three orbits, $\Gamma_1$,
$\Gamma_2$ and $\Gamma_3$, by use of the method M3a.} \label{fig1}
\end{figure}

\begin{figure*}[ht]
\includegraphics[scale=0.85]{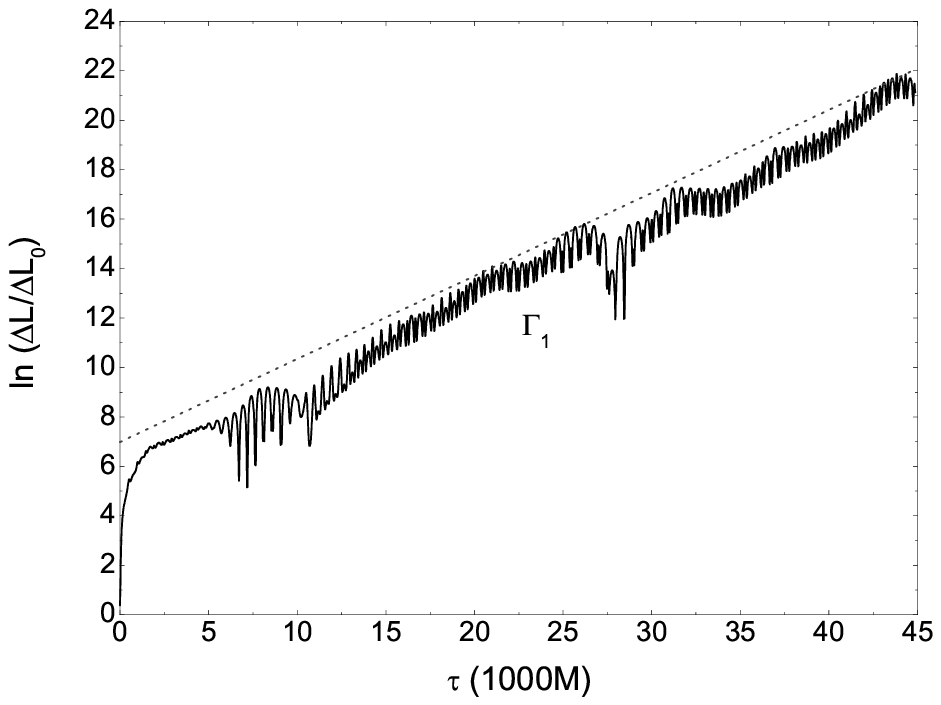}% Here is how to import EPS art
\includegraphics[scale=0.85]{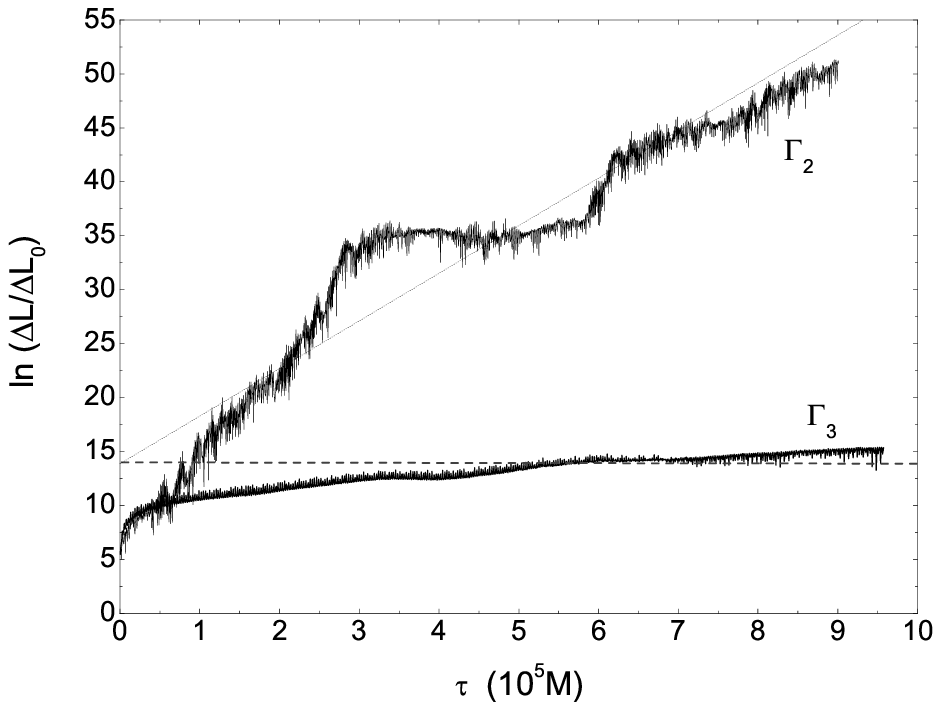} \caption{Invariant LEs as
plots of $\ln [\Delta L(\tau)/\Delta L(0)]$ vs $\tau$, based on
the method M3b. Curve $\Gamma_{1}$, with Lyapunov time
$\hat{\tau}_{\lambda}=2991M=10.9T_{o}$, nearly consists with the
lower line of Fig. 4 in Ref. [6]. The integration time is
$t=46000M$ for $\Gamma_{1}$, while $t=10^{6}M$ for $\Gamma_{2}$ or
$\Gamma_{3}$. } \label{fig2}
\end{figure*}

\begin{figure}[h]
\includegraphics[scale=0.85]{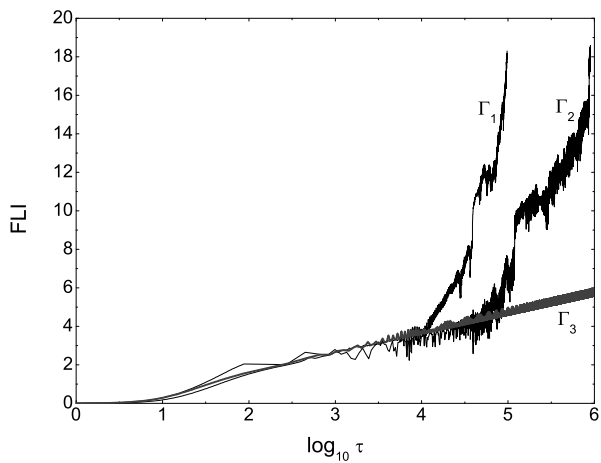}% Here is how to import EPS art
\caption{The evolution of invariant FLIs with proper time.}
\label{fig3}
\end{figure}

Now we conclude several problems that appear in Ref. [4]. (1) The
initial distance problem. An important point to note is that
relatively large and small initial separations are not permitted
when M2 is used. For a machine double-precision environment with
an order of $10^{-16}$, the starting distance $d(0)$ with a
magnitude of $10^{-8}$ is viewed as the best choice [10]. However,
Ref. [4] used $d(0)=10^{-10}$ that must give rise to the
overestimation of LEs if integration time is long enough. The
rescaling that brings roundoff errors can certainly have an effect
on LEs, but the initial distance is more important to affect LEs
than the rescaling. (2) The integration time problem. In general,
it is not true for the authors of Ref. [4] to declare the absence
of chaos in compact binaries by finding no stabilizing values of
$\ln \chi(t)$ only within a time span of a lower limit on the
Lyapunov time, $t_{\lambda}=1/\lambda$, with many times greater
than the typical inspiral time. Perhaps the authors considered
that chaos after the inspiral time scale will not affect the
dynamics of coalescing compact binaries. It is correct. However,
for M2a it usually takes many and many times greater than the
Lyapunov time (rather than the inspiral time) for $\ln \chi(t)$ to
approach to a certain stabilizing value. For instance, Ref. [11]
found that $\chi(t)$ of an orbit in 3-dimensional systems seems to
have been stabilized to a value near 0.0005 up to $t=220000$ and
then abruptly jumps to a value around 0.01 up to $t=1600000$ (see
Fig. 10a in Ref. [11]). That is to say, it is completely
impossible to arrive at the reliable value about 0.01 of LE when
the orbit is integrated to the Lyapunov time, 100. In sum, the
orbits of compact binaries must be integrated numerically for
sufficiently long times, otherwise there are unreliable LEs.
Unfortunately, coalescence does no longer give a chance to
numerical integration. As an illustration, for the conservative
system in which gravitational radiation is turned off, the
coalescence is not a consequence of energy loss but just that
these chaotic orbits happen to veer too close at some stage and
merge. It should be possible in principle to find pairs that
execute enough orbits that they do not coalesce before a lengthy
integration has been performed. (3) The coordinate gauge
invariance problem. There is a long history of the problem using
LEs  reliably in a curved space. In the mixmaster cosmology  there
was a long standing debate that the LEs were zero therefore there
was no chaos [12-15]. This was a wrong conclusion and was an
artifact of the spacetime slicing. Many independent groups have
been engaged to this field. For example, Imponente and Montain
[16] gave an invariant treatment of LEs by projecting a geodesic
deviation vector for the Jacobi metric on an orthogonal tetradic
basis so that they could successfully gain an insight into the
dynamics of the mixmaster cosmology. So did  Motter [17], who
addressed directly the issue of the invariance of LEs. The
invariant LEs in these works are mainly focused on the time
evolution of the gravitational field itself. However, relativistic
compact binaries are attributed to the geodesic or nongeodesic
motion of particles in a given gravitational field. Ref. [3] used
fractal basin boundary methods to detect chaos in black hole
pairs. It should be mentioned that the original conclusion that
there is  chaos in spinning binaries was made using a coordinate
invariant approach. There is no ambiguity in that approach. But
the fractals can't tell one more details of the dynamical
features, such as the timescale for chaos to set in. Thus, it is
fair to try to find a good invariant version of the LE. There are
several points regarding this. From the physical point of view, it
is questionable that the LEs, based on coordinate time $t$ and the
Cartesian distance $d(t)$ between two nearby trajectories in the
phase space of 12 dimensions, are used to discuss the dynamics of
this system in Ref. [4]. Compact binaries are such strong fields
that general relativistic effects become very apparent. On the
other hand, general relativity admits a free choice of space and
time coordinates so that the spacetime coordinates usually play
book-keeping only for events. Therefore, physical observable
quantities, such as the distance and the time, should be defined
as proper quantities instead of coordinate quantities. This is a
basic point from the theory of observation in general relativity.
Following this idea, Ref. [18] used proper time $\tau$ of an
``observer" and a proper configuration space distance $\Delta L
(\tau)$ between the observer and his ``neighbor" particles to
construct an \emph{invariant} LE (M3): $ \lambda = \lim_{\tau
\rightarrow \infty}\chi(\tau),$  where $\chi(\tau) =(1/\tau) \ln
[\Delta L(\tau)/\Delta L(0)].$  $\Delta L (\tau)= (h_{\alpha
\beta} \Delta x^{\alpha}\Delta x^{\beta})^{1/2}$, with the space
projection operator of the observer $h^{\alpha\beta}=
g^{\alpha\beta}+U^{\alpha}U^{\beta}$, and $\Delta x^{\beta}$ being
the deviation vector from the observer to the neighbor. Here
$g^{\alpha\beta}$ and $U^{\alpha}$ stand for metric tensor and
4-velocity of the observer, respectively. In practice, M3 is no
other than a direct modified and refined version of M2. Naturally,
M2a and M2b are corresponded to M3a and M3b. As an illustration,
the coordinate time $t$ still remains of a common time variable in
the equations of motion for the two particles, while the proper
time $\tau$ is from integration of the equation about $d\tau/dt$.
For the special case of $\tau\sim\ln t$, M3 fails to work well. In
spite of that, we do believe that M3 will be very useful and
simple to investigate spinning compact binaries, because it is in
the coordinate time $t$ that the equations of motion for the
compact binaries have been given by Ref. [2], and $\tau$ and $t$
have no the approximately logarithmic relation at all. (4) The
power spectra problem. The authors of Ref. [4] did not find
chaotic behavior in terms of the power spectra. This is because
the power spectra are difficult to distinguish among complicated
periodic orbits, quasi-periodic orbits and weakly chaotic orbits.
Usually the power spectra are not recommended to be a criterion
for evaluating chaos.

One main aim of the present paper is to re-review the results of
Ref. [4], as has been stated above. The other is more important to
use the covariant chaos indicator M3 (M3a and M3b) to investigate
spinning compact binaries so that we take the opportunity to
examine the related results in some references [4-6]. Considering
the slow convergence of LEs and the possible coalescence of two
stars, we suggest adopting a sensitive tool for detecting
chaos---the \emph{invariant} FLI of two nearby trajectories in a
curved spacetime [19]: $
 FLI(\tau) = \log_{10}[\Delta L(\tau)/\Delta
L(0)].$ The related details and applications of FLIs can be seen
in Refs. [19-23]. It stretches exponentially with (proper) time
for the chaotic orbit, but grows linearly with time in the regular
case. Throughout the work we use units $c=G=1$ and the signature
of a metric as $(-,+,+,+)$, and let Greek subscripts run from 0 to
3 and Latin indexes from 1 to 3.

In compact binaries, the evolution equations about the relative
position $\mbox{\boldmath$x$}$ and velocity $\mbox{\boldmath$v$}$
for body 1 relative to body 2 at 2PN order in harmonic coordinates
are $\ddot{\mbox{\boldmath$x$}}=\mbox{\boldmath$a$}_{N}
+\mbox{\boldmath$a$}_{1PN}
+\mbox{\boldmath$a$}_{1.5SO}+\mbox{\boldmath$a$}_{2PN}
+\mbox{\boldmath$a$}_{2SS}.$ The numbers and the letters denote
the order of the PN expansion and type of the contributions to the
relative acceleration, respectively. The two spins precess by $
\dot{\mbox{\boldmath$S$}}_{\imath}=\mbox{\boldmath$\Omega$}_{\imath}
\times \mbox{\boldmath$S$}_{\imath} ~(\imath=1,2).$ Their explicit
forms can be seen in Ref. [2]. Now let $m_{\imath}$ be mass of
body $\imath$, and the total mass $M=m_{1}+m_{2}$. In addition, we
specify $(\mbox{\boldmath$y$}_{\imath},
\mbox{\boldmath$v$}_{\imath})$ as position and velocity of each
body in the center-of-mass (CM) frame. The relations among three
coordinates $\mbox{\boldmath$y$}_{1}$, $\mbox{\boldmath$y$}_{2}$
and $\mbox{\boldmath$x$}$ at 2PN order are $
 \mbox{\boldmath$y$}_{1} = (m_2/M)\mbox{\boldmath$x$}
+Y_{1PN}(\mbox{\boldmath$x$},\mbox{\boldmath$v$})
+Y_{1.5}(\mbox{\boldmath$S$}_1,\mbox{\boldmath$S$}_2)
+Y_{2PN}(\mbox{\boldmath$x$},\mbox{\boldmath$v$})$ and
$\mbox{\boldmath$y$}_{2}= -(m_1/M)\mbox{\boldmath$x$}
+Y_{1PN}(\mbox{\boldmath$x$},\mbox{\boldmath$v$})
+Y_{1.5}(\mbox{\boldmath$S$}_1,\mbox{\boldmath$S$}_2)
+Y_{2PN}(\mbox{\boldmath$x$},\mbox{\boldmath$v$})$  [24]. Here,
the 1PN and 2PN terms can be found in Ref. [25], while the 1.5
order term is given by Ref. [26]. On the other hand, the proper
time $\tau$ of body 1 in the CM frame satisfies the equation$
d\tau/dt=[-(g_{00}+2g_{0i}v^{i}_{1}
+g_{ij}v^{i}_{1}v^{j}_{1})]^{1/2}.$  $g_{\alpha\beta}$, as a
function of $(\mbox{\boldmath$y$}_{1},\mbox{\boldmath$y$}_{2};
\mbox{\boldmath$v$}_{1},\mbox{\boldmath$v$}_{2})$, is the 2PN
metric tensor at body 1. Each metric component is made of the
related potentials at the location of body 1, and each potential
is the sum of the non-spin piece and of the spin part. The
non-spin part is presented by Ref. [27], and the spin piece is
listed in Ref. [26]. See also Ref. [28] that contains the sum of
the two parts. As an illustration, the 2.5 order terms in the
references are dropped. This physically corresponds to dropping
dissipative terms.

Clearly, the coordinate time $t$ plays an important role in
connection with the motion of body 1 and of body 2, and the
relative motion between the two bodies, but proper time does not
since it differs for each of three motions. This gives us a good
chance to apply M3 and the metric $g_{\alpha\beta}$ to study the
dynamics of orbits around body 1 in the  CM frame [29]. The
implementation is described briefly. We integrate the equations
(7) of the relative motion, the spin equations (8) and the proper
time equation (10) numerically together by using a fifth-order
Runge-Kutta-Fehlberg algorithm with an adaptive coordinate time
step. At once, we can get $\mbox{\boldmath$S$}_1$,
$\mbox{\boldmath$S$}_2$, $\mbox{\boldmath$x$}$,
$\mbox{\boldmath$v$}$, and $\tau$, at coordinate time $t$. Then
$(\mbox{\boldmath$y$}_{1},\mbox{\boldmath$y$}_{2};
\mbox{\boldmath$v$}_{1},\mbox{\boldmath$v$}_{2})$ are determined,
and body 1 has its 4-velocity $
\mbox{\boldmath$U$}=(\frac{dt}{d\tau}, v^{1}_{1}\frac{dt}{d\tau},
v^{2}_{1}\frac{dt}{d\tau}, v^{3}_{1}\frac{dt}{d\tau}).$ Now body 1
is chosen as an observer, who can measure the proper distance
$\Delta L$ to his neighboring orbit. Note, the neighboring orbit
is not the orbit at which body 2 stays, and is from an orbit
nearby body 1, caused by a slight separation of the relative
position. In a word, numerical integration is to carry out in the
relative coordinate system, but the relativistic dynamics is to
investigate in the CM frame and whether chaos or not is measured
by body 1. This is entirely different from the treatment of other
references, where the Newtonian dynamical methods are used to
consider the relative motion in spinning compact binaries.

Let us re-calculate the LEs of three orbits that had been studied
in Ref. [6]. The related initial conditions and parameters of the
orbits are listed here. Orbit $\Gamma_1$: phase space variables
$(\mbox{\boldmath$x$},\mbox{\boldmath$v$})=(5.5M,0,0,0,0.4,0)$,
mass ratio $\beta=m_2/m_1=1/3$, spin magnitudes
$S_{\imath}=m^{2}_{\imath}$, and spin directions
$\theta_{1}=\pi/2$ and $\theta_{2}=\pi/6$.  $\Gamma_2$:
$(\mbox{\boldmath$x$},\mbox{\boldmath$v$})=(5.0M,0,0,0,0.399,0)$,
$\beta=1$, $S_{\imath}=m^{2}_{\imath}$, $\theta_{1}=38^{\circ}$,
and $\theta_{2}=70^{\circ}$. $\Gamma_3$ is the same as $\Gamma_2$
but only 0.428 replaces 0.399. In addition, let only the first
component $x$ of the initial relative position of each trajectory
have a very small deviation, $\Delta x=10^{-8}M$, then we get its
corresponding neighboring orbit. Following M3a, we draw plots of
$\log_{10}\chi(\tau)$ vs $\log_{10}\tau$ about the LEs of the
three trajectories in Fig. 1. They all drop before proper time
$\tau$ spans $10^{6}M\approx 3636T_{o}$ ($T_{o}=275M$, the average
period of orbit $\Gamma_1$). For $\Gamma_1$, the LE time looks to
get a reliable value, $\tau_{\lambda}=4675M=17.0T_{o}$. It is
larger than the value $t_{\lambda}=3080M=11.2T_{o}$ given in Ref.
[6]. Perhaps one addresses a question whether $\chi$ can still
remain the value of $1/\tau_{\lambda}$ if numerical integration
continues. We have no way to answer it since numerical integration
has to end after $t=1.37\times 10^{6}M$ (or $\tau=1335980M$), when
the two objects coalesce. Above all, neither $\Gamma_2$ nor
$\Gamma_3$ has any acceptable stabilizing value when integration
time $t$ reaches $10^{8}M$. Additionally, we did not find any
difference from these results by making several tests with
different renormalization time steps. This seems to show that the
results in Ref. [4] are reasonable. However, the case is
completely different when M3b is used. Seen from the calculations
including the rescaling, M3b is almost the same as M3a, but only a
slight difference between them lies in plotting $\ln [\Delta
L(\tau)/\Delta L(0)]$ vs $\tau$ instead of plotting
$\log_{10}\chi(\tau)$ vs $\log_{10}\tau$. Another point to note is
that M2b (\emph{rather than M3b}) without rescaling was used in
Ref. [7], but our M3b employs rescaling. In addition, there is a
difference that the authors of Ref. [7] use the Hamiltonian
formulation in ADM coordinates and not the Lagrangian formulation
in harmonic coordinates. Although the two approaches are
approximately related, they are not exactly equal. For instance,
the constants of motion  are exactly conserved in the Hamiltonian
formulation, while they are approximately in the Lagrangian
formulation. Ref. [7] also works to 3PN order. As shown in Fig. 2,
it takes no long enough time to explicitly see the presence of
positive slopes of the fit lines for $\Gamma_1$ and $\Gamma_2$,
but to do that of about zero slope of the fit line for $\Gamma_3$.
This means chaos of $\Gamma_1$ and $\Gamma_2$, while the
regularity of $\Gamma_3$. It is what can be seen in Refs. [5,6].
It is sufficiently argued that the LEs converge much faster for
M3b than for M3a. As an illustration, $\tau_{\lambda}$ is more
reliable than $\hat{\tau}_{\lambda}$. This is because the longer
numerical integration becomes, the more accurate the values of LEs
are. In fact, the fit slope (its inverse being $4680M$) of
$\Gamma_{1}$ in Fig. 2 is very close to the LE of $\Gamma_{1}$ in
Fig. 1 when integration times are the same. Now, we can say quite
plainly that a reliable conclusion is that there is chaos in the
2PN system. So can the authors of Ref. [7], who have already
confirmed the existence of chaos in the 2PN Hamiltonian
formulation through positive LEs. As mentioned above, although the
LEs converge much faster for M3b than for M3a, long integration
times are still needed to get reliable values of LEs even if M3b
is considered. Noting this, we recommend to use a quicker
indicator, the invariant FLI given by Eq. (6). Its algorithm can
be found in Ref. [19]. Fig. 3 displays that FLIs of $\Gamma_1$ and
$\Gamma_2$ increase exponentially with $\log_{10}\tau$, but that
of $\Gamma_3$ does algebraically. Thus $\Gamma_1$ and $\Gamma_2$
are chaotic, (chaos of $\Gamma_1$ is much stronger than that of
$\Gamma_2$) but $\Gamma_3$ becomes ordered. It is worth
emphasizing that the three orbits can be distinguished clearly in
practice when proper time adds up to $10^{5}M$. Consequently, the
onset of chaos in the 2PN Lagrangian approximation  is proved
again through different ratios of FLIs varying with time.

The summary is included as follows. For conceptual clarity, it is
necessary to apply chaos indicators independent of the choice of
coordinate gauge to analyze the dynamics of relativistic
gravitational systems. Since coordinate time is a good medium in
connect with the mass centered motions of both body 1 and body 2,
and the relative motion in spinning compact binaries, we think
that M3, as an invariant indicator, is a good tool to study these
systems. Using M3, we estimate the LEs on the  mass centered
motion of body 1 rather than on the relative motion considered by
other references. We find that the orbits must be calculated for
long enough times in order to get stabilizing limit values as
reliable LEs for the case of comparable mass compact binaries. On
the other hand, we track that the exact source of both the failure
of Ref. [4] and the success of Refs. [5,6] in the computation of
LEs does not stem from the rescaling, but is based on two slightly
different treatments of LEs, M3a and M3b. At most cases, the LEs
converge much faster for M3b than for M3a. However, coalescence of
the black holes makes it impossible in some cases to have enough
numerical integration. This shows that M3a is no longer a suitable
indicator to quantify chaos in spinning compact binaries.
Additionally, it should be noted that although it is rather easier
to get the LEs for M3b than for M3a, long integration times are
still needed to get reliable values of LEs when M3b is adopted. In
this sense, the invariant FLI in a curved spacetime is a very fast
and valid technique to detect chaos from order. Still, a reliable
conclusion is that there is chaos in the conservative 2PN
Lagrangian system. Of course, the 2PN approximation is so poor
that there has been left an open question whether real binary
systems with better approximations exhibit chaos [30]. Saying this
another way, now one does not say that chaos can be ruled out in
real binaries. In future, we will discuss a wider application of
the FLI in detailedly investigating the dynamics of spinning
compact binaries.

%\begin{acknowledgments}
We would like to thank both the referee and F.A. Rasio for their
honest comments and significant suggestions. We are also grateful
to Professor Tian-Yi Huang of Nanjing University for his helpful
discussion. This research is supported by the Natural Science
Foundation of China under Contract No. 10563001. It is also
supported by the Science Foundation of Jiangxi Province (0612034),
the Science Foundation of Jiangxi Education Bureau (200655), and
the Program for Innovative Research Team of Nanchang University.
%\end{acknowledgments}


\begin{thebibliography}{}
\bibitem{24} The relation up to 2PN order is used to get the
proper distance $\Delta L$ with the same accuracy. Of course, the
terms higher than 2  order in $\mbox{\boldmath$v$}_1$ and
$\mbox{\boldmath$v$}_2$ are turned off.
\bibitem{29} As a reason to do this, the metric is easily given in
the frame. In addition, the LE by M3 does not depend on a
spacetime coordinate system. This implies that the LE of body 1 in
the frame is equivalent to one of the relative motion orbit.
\end{thebibliography}
\end{document}